\documentclass[aps,prl,twocolumn,groupedaddress]{revtex4}

\usepackage{graphicx}

\begin{document}


\title{Coexistence of antiferromagnetism and superconductivity in CeRhIn$_5$ under high pressure and magnetic field}


\author{G. Knebel}
\email{gknebel@cea.fr}
\affiliation{D\'{e}partement de la Recherche Fondamentale sur la Mati\`{e}re Condens\'{e}e, SPSMS, CEA Grenoble, 17 rue des Martyrs, 38054 Grenoble Cedex 9, France}
\author{D. Aoki}
\altaffiliation[Permanent address: ]{IMR, Tohoku University, Ibaraki, Japan}
\author{D. Braithwaite}
\affiliation{D\'{e}partement de la Recherche Fondamentale sur la Mati\`{e}re Condens\'{e}e, SPSMS, CEA Grenoble, 17 rue des Martyrs, 38054 Grenoble Cedex 9, France}
\author{B. Salce}
\affiliation{D\'{e}partement de la Recherche Fondamentale sur la Mati\`{e}re Condens\'{e}e, SPSMS, CEA Grenoble, 17 rue des Martyrs, 38054 Grenoble Cedex 9, France}
\author{J. Flouquet}
\affiliation{D\'{e}partement de la Recherche Fondamentale sur la Mati\`{e}re Condens\'{e}e, SPSMS, CEA Grenoble, 17 rue des Martyrs, 38054 Grenoble Cedex 9, France}



\date{\today}

\begin{abstract}
We report on detailed ac calorimetry measurements under high pressure and magnetic field of CeRhIn$_5$. Under hydrostatic pressure the antiferromagnetic order vanishes near $p_c^\star=2$~GPa due to a first order transition. Superconductivity is found for pressures above 1.5~GPa inside the magnetic ordered phase. However, the superconductivity differ from the pure homogeneous superconducting ground state above 2~GPa. The application of an external magnetic field $H \parallel ab$ induces a transition inside the superconducting state above $p_c^\star$ which is strongly related to the re-entrance of the antiferromagnetism with field. This field-induced supplementary state vanishes above the quantum critical point in this system. The analogy to CeCoIn$_5$ is discussed. 
\end{abstract}


\maketitle


The discovery of the cerium heavy fermion Ce$M$In$_5$ $(M={\rm Co, Ir, Rh})$ systems opened a new route to investigate the interplay of magnetic fluctuations, unconventional superconductivity (SC) and antiferromagnetic order (AF) in strongly correlated electron systems in the vicinity of a quantum phase transition (QPT). While CeCoIn$_5$ and CeIrIn$_5$ are unconventional superconductors at ambient pressure ($p$)with most probably $d$-wave symmetry below $T_c = 2.2$~K and $T_c = 0.4$~K and suited very close to a QPT, CeRhIn$_5$ offers the possibility to tune a heavy fermion system from AF to SC as function of pressure \cite{Hegger2000,Petrovic2001,Petrovic2001b}. At ambient pressure CeRhIn$_5$ orders antiferromagnetically below the N\'eel temperature $T_N = 3.8$~K in an incommensurate helical structure with propagation vector $Q = (1/2, 1/2, 0.297)$ and a staggered moment of about 0.8~$\mu_B$ \cite{Bao2003}. 
Neutron scattering experiments under pressure showed that this magnetic structure is almost unchanged under pressure up to the highest investigated pressure of 1.7~GPa \cite{Llobet2004}. SC has been reported for pressures above 0.9~GPa on the basis of resistivity measurements. Nuclear-quadrupole resonance (NQR) measurements in combination with ac susceptibility have shown that SC and AF coexist on a microscopic scale in the pressure range from $p= 1.6 - 1.75$~GPa \cite{Kawasaki2003}. The spin-lattice relaxation rate shows at low temperature an unexpected $1/T_1 \propto T$ dependence which gives evidence for a gap-less nature of the low-lying excitations. A pure superconducting ground-state with line nodes in the gap is attained for $p>2$~GPa and $1/T_1 \propto T^3$ has been reported \cite{Mito2001}. Previous detailed specific heat measurements under hydrostatic pressure give evidence that the magnetic transition disappears by a first order transition at $p_c^\star \approx 1.95$~GPa \cite{Knebel2004}. Measurements of quantum oscillations under pressure show the appearance of new frequencies for $p> 2.34$~GPa, pointing to a reconstruction of the Fermi surface near a quantum critical point (QCP) \cite{Shishido2005}.

The application of an external magnetic field $H$ offers a second parameter to influence the competition between AF and SC. Recently it was shown that a magnetic QCP can be achieved in CeCoIn$_5$ by the application of a magnetic field of the order of the upper critical field $H_{c2} (0)$ \cite{Paglione2003,Bianchi2003c}. $H_{c2}(T)$ in this compound is strongly Pauli limited and the transition becomes first order at low temperatures \cite{Bianchi2002,Takeuchi2002,Tayama2002}. At high magnetic field near $H_{c2} (0)$ a new superconducting phase was observed which may be the first example of a Fulde-Ferrel-Ovchinnikov-Larkin (FFLO) state in an inorganic superconductor \cite{Bianchi2003b,Radovan2003,Kaguyanagi2005}. However, this new state cannot be interpreted as a stack of spatially homogeneous superconducting and normal regions \cite{Mitrovic2005}.

The high magnetic field phase diagram of CeRhIn$_5$ at ambient pressure has been studied by specific heat \cite{Cornelius2001}, magnetization \cite{Cornelius2000,Takeuchi2001}, and thermal expansion measurements \cite {Correa2005}. For fields in the $ab$-plane, $T_N$ is slightly increasing with increasing field up to 3.9~K. The magnetization in the $ab$-plane shows a small step-like increase near $H_m^{\star} = 2$~T at low temperatures, due to a tiny change in the magnetic structure presumably associated with a reorientation of the helix and/or a change from an incommensurate to a commensurate structure. Consequently for $H > H_m^{\star}$ a second magnetic transition below $T_N$ appears in temperature dependent measurements at $T_{N2} \approx 3.5$~K. $T_{N2}$ is also slightly increasing with field. A transition to a spin-polarized state is observed for magnetic fields above $H_m\approx 50$~T \cite{Takeuchi2001}. For fields in the $c$-direction no second phase under field appears. 

In this letter we report on ac calorimetric measurements under high pressure and under magnetic field in CeRhIn$_5$ for temperatures above 0.5~K in the pressure range 1.6~GPa $<p<$ 2.73~GPa. For $p<p_c^\star \approx 1.95$~GPa we observed very tiny superconducting specific heat anomalies in the coexistence regime of the phase diagram in zero field. Above $p_c^\star$ the application of a magnetic field in the $ab$ plane leads to a re-entrance of the magnetic order and coexistence of AF and SC is observed below $H_{c2}$. The reentrant phase vanishes at the critical pressure $p_c$ where the N\'{e}el temperature $T_N$ collapses under field. 

The pressure cell used is the same as in the previous experiment \cite{Knebel2004}. Details of the ac calorimetry are discussed elsewhere \cite{Derr2005}. Heating is realized in this low temperature experiment by a laser diode which can be tuned in frequency and power. The frequency was chosen to be slightly above the cut-off $\nu_c \approx 300$~Hz at the lowest temperature independent of pressure. The specific heat can be estimated by $C_{ac} \propto - S_{th} sin\theta / V_{th} 2\pi \nu$ where $S_{th}$ corresponds to the thermoelectric power of an AuFe/Au thermocouple which is soldered on the sample, $V_{th}$ and $\theta$ are the measured voltage and phase of the thermocouple signal. The measurements were performed in a $^3$He cryostat where a magnetic field of 7.5~T could be applied in the $ab$ plane of the single crystals.

 \begin{figure}
 \includegraphics[width=0.38\textwidth]{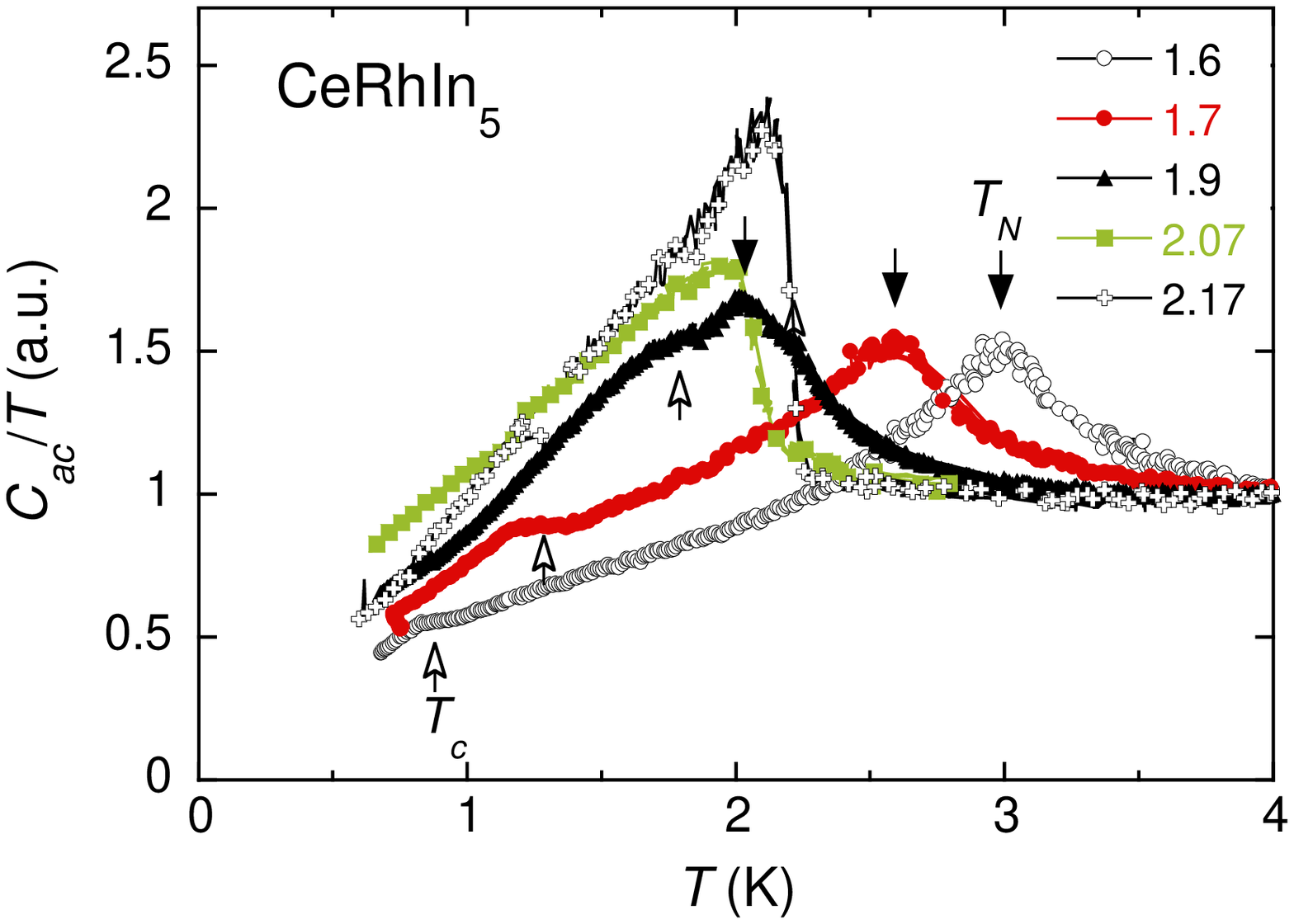}%
 \caption{\label{Cac_T_fig1}Temperature dependence of the ac specific heat divided by temperature for different pressures. Open Arrows indicate the superconducting transition temperatures $T_c$, closed arrows the N\'{e}el temperature. Below $p_c^\star \approx 1.95$~GPa the superconducting anomaly in the specific heat is very small. For $p$ slightly above $p_c^\star$ a nice superconducting anomaly appears.(Data are normalized to 1 at 4~K.)}
 \end{figure}
 
  \begin{figure}
 \includegraphics[width=0.38\textwidth]{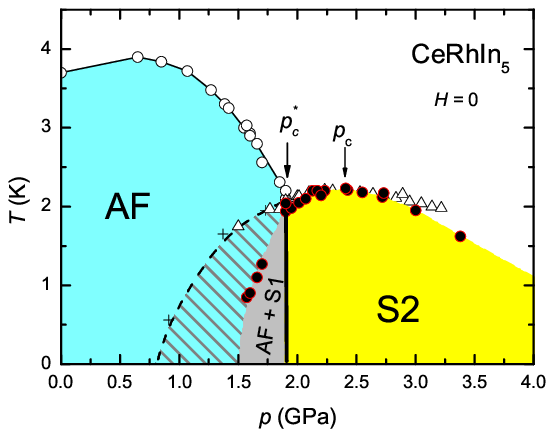}%
 \caption{\label{p-T_phase-diagram}$p-T$ phase diagram of CeRhIn$_5$ from specific heat (circles), susceptibility (triangles) and resistivity measurements (crosses, from Ref.\onlinecite{Llobet2004}). Below 1.5~GPa the ground state is an AF, the hatched area indicates an inhomogeneous superconducting state, AF+S1 corresponds to the region where SC appears in the specific heat experiment in the magnetically ordered state below $T_N$. However, the superconducting order parameter is different from the purely superconducting state S2 realized above $p_c^\star$. The vertical line marks the first order transition from AF+S1 to S2. $p_c$ corresponds to the critical pressure, where the reconstruction of the Fermi surface appears \cite{Shishido2005}.  }
 \end{figure}

Figure \ref{Cac_T_fig1} shows the specific heat divided by temperature in zero field in the pressure range from $1.6-2.17$~GPa. In contrast to the previous measurements which had been limited to temperatures above 1.4~K \cite{Knebel2004} in the new measurements two specific heat anomalies appear in the pressure range from 1.6-1.9~GPa. However, the lower temperature transition, which corresponds to a superconducting transition, is very tiny for $p<p_c^\star \approx 1.95$~GPa. In contrast to this, the measurement clearly shows a sharp and very large superconducting anomaly at $p=2.17$~GPa indicating a pure superconducting ground state. 

 \begin{figure}
 \includegraphics[width=0.35\textwidth]{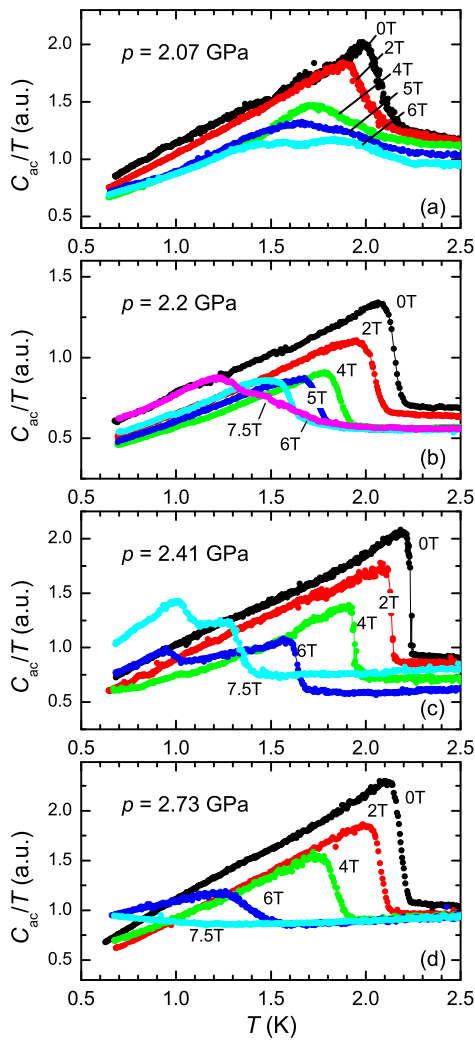}%
 \caption{\label{Cac_field}Specific heat $vs.$ temperature of CeRhIn$_5$ at different pressures for various magnetic field $H \parallel ab$. }
 \end{figure}

The pressure-temperature phase diagram of CeRhIn$_5$ in zero field is summarized in Fig. \ref{p-T_phase-diagram}. The low temperature specific heat measurements clearly show the coexistence of antiferromagnetism and superconductivity in the pressure range from 1.6-1.9~GPa. The observed superconducting anomalies in the resistivity $T_c^\rho$and susceptibility measurements $T_c^\chi$ are obviously due to inhomogeneities in the low pressure regime ($p< 1.5$~GPa) and in the temperature interval $T_c^\rho >T> T_c^C$.  The $T_c$ determined by NQR measurements agrees perfectly with the specific heat results $T_c^C$ \cite{Kawasaki2003}.
A first order QPT emerges at $p_c^\star=1.95$~GPa, where the superconducting transition temperature $T_c = T_N$. A linear extrapolation of $T_N$ to zero temperature indicates a QCP for $p_c \approx 2.5$~GPa, very close to the pressure where the reconstruction of the Fermi surface was observed in the high field dHvA experiments. However, for $p>p_c^\star$ no antiferromagnetic order appears in zero field. If $T_c > T_N$ the ground state in zero field is purely superconducting with most probably $d$-wave symmetry. 
The opening of the superconducting gap on large parts of the main Fermi surface suppresses the possibility of magnetic ordering. This  situation is quite different from the behavior below $p_c^\star$, where only very tiny superconducting anomalies $T_c <T_N$  are observed. This tiny anomaly indicates an important residual density of states at the Fermi surface in this pressure regime. However, from our experiment we cannot definitely conclude if a homogeneous gap-less superconducting state is formed which coexists with magnetic order or if a
phase separation with AF and SC volume fraction appears, as observed in CeIn$_3$ e.g.\cite{Kawasaki2004}. However, the observation of only two peaks at low temperatures in the $^{115}$In NQR spectra due to the onset of AF at 1.6~GPa and 1.75~GPa seems to exclude an inhomogeneous state \cite{Kawasaki2003,Mito2003}. Very recently phase separation has been observed at 1.95~GPa, in agreement with the first order transition at his pressure \cite{Yashima2005}.

\begin{figure}
 \includegraphics[width=0.35\textwidth]{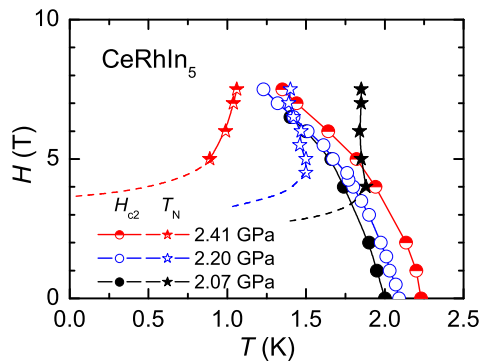}%
 \caption{\label{parameterplot}$H-T$ phase diagram for different pressures. Circles give the upper critical field $H_{c2}$, stars correspond to a field induced magnetic transition. The dashed lines indicate an extrapolation of $T_N$ to $H=0$ and give the low field limit for the detection of the magnetic state, solid lines for $H_{c2}$ are guides for the eye. }
 \end{figure}

In the following we want to discuss the effect of a magnetic field on the superconducting state above $p_c^\star$. We performed specific heat measurements at fixed field $H$. Figure \ref{Cac_field} shows the specific heat for various fields $H\parallel ab$ at different pressures above $p_c^\star$. The superconducting transition in zero field at 2.07~GPa is still rather broad. Under magnetic field the transition broadens further, but for $H \gtrsim 4$~T an additional shoulder develops in the transition. For 6~T clearly two maxima appear, indicating two different phase transitions. At 2.2~GPa, the total width of the transition is less than 0.1~K. Again, for fields higher than 4~T the transition gets significantly broader, indicating a second transition under  field. The appearance of a second phase becomes very clear at 2.41~GPa with two nicely defined transitions. For higher pressures ($p=2.73$~GPa $<p_c$) we observed only the superconducting phase. 

The temperature dependence of the upper critical field and the temperature of the appearance of the field-induced magnetic state are plotted in Fig.~\ref{parameterplot} for different pressures. Let us first discuss the superconducting parameters. Under consideration of the most simple model, weak coupling in the clean limit, the superconducting coherence length $\xi_0$ and an average Fermi velocity $v_F$ can be estimated from the initial slope $H_{c2}'=dH_{c2}/dT$. For $p=2.07$~GPa we find $H_{c2}'= 20$~T/K, near 2.41~GPa $H_{c2}'$ increases to 33~T/K and decreases to 25~T/K for 2.73~GPa. From this and the maximal $T_c$ it follows directly, that the superconducting coherence length is smallest near $p \approx 2.5$~GPa and we find $\xi_0 = 28$~\AA\ for 2.41~GPa, comparable to data published in Ref.\onlinecite{Muramatsu2001}. The Fermi velocity is  about 4550~m/s for this pressure. This indicates that the effective mass $m^\star = \hbar k_F/ v_F$  of the quasi-particles has its maximum near 2.5~GPa, pointing to a QCP at this pressure. Furthermore the relative size of the superconducting jump at $T_c$ is highest near 2.5~GPa (see also Ref.\onlinecite{Knebel2004}. Like in CeCoIn$_5$ the upper critical field in CeRhIn$_5$ is strongly Pauli limited. The upper critical field due to the orbital effect can be estimated from the initial slope $H_{c2}^{orb} = 0.7 H_{c2}'T_c = 51$~T for 2.41~GPa. The Pauli limiting, given by $H_{c2}^P = 2.25 T_c$~T/K$= 5$~T is about ten times lower for this pressure. 

Now we discuss the appearance of a field induced antiferromagnetic state in CeRhIn$_5$ for $p > p_c$. For $p=2.07$~GPa we did not find evidence for magnetic order in zero field (see Fig.\ref{Cac_field}a) but two anomalies appear for $H>4$~T and $T_N$ is slightly above $T_c$. As for the low pressure AF phase ($p< 1.6$~GPa) the phase transition above $T_c$ is almost field independent. An extrapolation to lower fields give $T_N=T_c \approx 1.9$~K for $H \approx 1$~T. At 2.2~GPa the broadening of the superconducting transition above 4~T indicates the appearance of the magnetic phase inside the superconducting regime. At $H\approx 6.3$~T, $T_N \approx T_c \approx 1.4$~K, and for higher field, $T_N (H) > T_c (H)$. Two well separated anomalies have been detected for $p = 2.41$~GPa above 5~T. Obviously, for this pressure a new phase transition appears inside the superconducting state as $T_N < T_c$ for all fields below 7.5~T and above 4~T which coexist with the superconducting state. Here we can extrapolate $T_N  \approx T_c \approx 1$~K for $H=8$~T. The situation is drastically different from the behavior below $p_c^\star$. At $p=2.73$~GPa, no magnetic transition appears in the temperature and field range of our measurements. If we extrapolate $T_N = T_c$ to zero temperature with pressure and field as implicit variables, the field will be very near the upper critical field of about 10~T for $p_c \approx 2.65 \pm 1.5$~GPa, not so far from the pressure of the maximum of $T_c$ and the maximum of the effective mass in zero field.

\begin{figure}
 \includegraphics[width=0.35\textwidth]{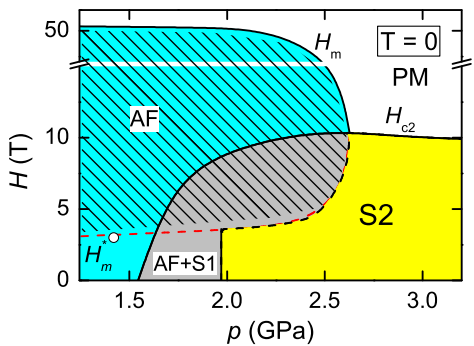}%
 \caption{\label{H-P_diagram}$H-p$ phase diagram of CeRhIn$_5$ at $T=0$. AF marks the pure antiferromagnetic phase, the hatched region corresponds to the regime where the magnetic structure is different from the incommensurate determined at $p=0$ and $H=0$. AF+S1 the coexistence regime of SC and AF, S2 represents the pure superconducting phase.  }
 \end{figure}

We want to emphasize that coexistence of AF and SC for $p>p^\star$ appears only above some critical field of the order $H\approx 4$~T. AF order can originate from the vortex core. The field must be sufficiently high that the spacing of the vortex lattice is small enough so that a magnetically coherent state can be formed.
Furthermore it is interesting to note that the field where the AF order above $p_c^\star$ appears is of the same order as the field $H_m^\star$ where the magnetic structure is slightly changing for $H \parallel ab$ at ambient pressure. There is strong evidence that this characteristic field is almost constant under pressure. We found by ac calorimetry for 1.42~GPa almost the same $H-T$ phase diagram as has been published in Ref.\onlinecite{Cornelius2001} (not shown) with $H_m^\star \approx 3.5$~T for $p=1.42$~GPa. However, detailed neutron scattering or NMR experiments under pressure are indispensable to resolve the nature of the magnetically ordered state inside the superconducting state above $p_c^\star$.  

To summarize we have plotted schematically in Fig.\ref{H-P_diagram} the $H-p$ phase diagram of CeRhIn$_5$ at $T=0$. Up to 1.5~GPa the ground state is purely antiferromagnetic. The magnetic properties are almost unchanged in comparison to $p=0$. There is a modification of the magnetic structure for $H_m^\star \gtrsim 2$~T. The saturation field $H_m$ is surely one order of magnitude higher than this field. From 1.5 - 2~GPa the ground-state is AF+S1, and the superconducting component differ from the pure superconducting phase S2 which appears above $p_c^\star \approx 2$~GPa due to the interplay with the magnetic structure. To modify the nature of the superconducting state S2 above $p_c^\star$  via the $H$ re-entrance of AF a minimal field $H\sim H_m^\star$ seems to be required. Furthermore close to $p_c$ for $p=p_c-\epsilon$, two sharp superconducting phase transitions occur above $H_m^\star$ suggesting two distinct order parameters; the superconducting matter of the low temperature - high magnetic field phase may be due to the unique situation of the proximity of both, magnetic and superconducting coherence lengths related respectively to the magnetic QPT and clean unconventional superconductivity. These experiments suggest strongly that the so-called FFLO state in CeCoIn$_5$ may be revisited. We want to emphasize that the first order separation between AF and S2 may be a quite general phenomenon at least for heavy fermion superconductors.

\begin{acknowledgments}
We want thank J.-P. Brison, K. Izawa, M. Houzet, V. Mineev, K. Miyake and M. Zhitomirsky for stimulating discussions. This work has been supported by IPMC, Grenoble. 
\end{acknowledgments}


\end{document}